%% file: preprint.tex
\documentclass{article}
\usepackage{amssymb}
\RequirePackage{filecontents}
\PassOptionsToPackage{dvipsnames}{xcolor}
\RequirePackage{comgipp}
\RequirePackage{amsfonts}
\RequirePackage{amsmath}
\RequirePackage{authblk}

\newcommand{\theReferenceText}{\fullcite{Ihle2020}}
\input{header.tex}
\input{authblk.tex}

\begin{document}
  \maketitle
  \thispagestyle{firststyle}
  \begin{abstract}
\input{abstract.tex}
  \end{abstract}
\input{mainmatter.tex}
\end{document}

%% file: header.tex
\usepackage[
backend=biber,
style=numeric,
firstinits=true,
url=false,
isbn=false,
safeinputenc,%
]{biblatex}

\usepackage{cleveref}
\usepackage{multirow}
\usepackage{booktabs}

\title{A First Step Towards Content Protecting Plagiarism Detection}

%% file: authblk.tex
\author[1]{Cornelius Ihle}
\author[2]{Moritz Schubotz}
\author[3]{Norman Meuschke}
\author[4]{Bela Gipp}
\affil[1]{Daimler AG \& Univ. of Wuppertal, Germany ({\{first.last\}@daimler.com})}
\affil[2]{FIZ-Karlsruhe, Germany (\{first.last\}@fiz-karlsruhe.de)}
\affil[3,4]{Universities of Wuppertal \& Konstanz, Germany (\{last\}@uni-wuppertal.de)}

%% file: abstract.tex
Plagiarism detection systems are essential tools for safeguarding academic and educational integrity.
However, today’s systems require disclosing the full content of the input documents and the document collection to which the input documents are compared.
Moreover, the systems are centralized and under the control of individual, typically commercial providers.
This situation raises procedural and legal concerns regarding the confidentiality of sensitive data, which can limit or prohibit the use of plagiarism detection services.
To eliminate these weaknesses of current systems, we seek to devise a plagiarism detection approach that does not require a centralized provider nor exposing any content as cleartext.
This paper presents the initial results of our research.
Specifically, we employ Private Set Intersection to devise a content-protecting variant of the citation-based similarity measure Bibliographic Coupling implemented in our plagiarism detection system HyPlag.
Our evaluation shows that the content-protecting method achieves the same detection effectiveness as the original method while making common attacks to disclose the protected content practically infeasible.
Our future work will extend this successful proof-of-concept by devising plagiarism detection methods that can analyze the entire content of documents without disclosing it as cleartext.

%% file: mainmatter.tex
\section{Introduction}
Plagiarism, i.e., the unacknowledged reuse of ideas or content, is a severe form of academic misconduct. 
Today, educational and research institutions, academic publishers, and funding agencies increasingly rely on plagiarism detection systems (PDS) to identify plagiarized content \cite{Weber-Wulff2019}. Typical PDS require users to submit input documents, which the systems then compare to a large, typically proprietary database of documents. The systems retrieve documents with similar content as the input document and highlight the similar content to support user inspection \cite{FoltynekMG19}.

Researchers and practitioners have criticized PDS for their poor detection accuracy and opaque computations \cite{Weber-Wulff2019}, as well as their centralized and nontransparent data management \cite[p.~72ff.]{WeberWulff2014}. In past research, we have addressed the first issue. We improved detection rates for heavily disguised instances of academic plagiarism by integrating the analysis of text-independent content elements, such as academic citations, images, and mathematical content with text-based detection methods into the hybrid plagiarism detection system HyPlag \cite{meuschke_hyplag:_2018}. 
In this paper, we focus on the second issue.

Disclosing the full content of input and comparison documents to a central service provider whose detection methods and data protection measures are nontransparent, inherently raises concerns regarding the security and confidentiality of sensitive data, e.g., in unpublished research grant proposals or research theses compiled in cooperation with companies. The mere disclosure of such content to a third party can violate non-disclosure agreements, as well as data protection and copyright laws \cite[p.~73f.]{WeberWulff2014}. Such legal concerns can limit the use of plagiarism detection systems. The general risk of data breaches further aggravates the problem. In Germany, many universities, therefore, prohibit the use of PDS entirely.

As we described in our vision paper \cite{ihle_privacy-preserving_2019}, we seek to address the weaknesses of current PDS by devising a blockchain-backed decentralized approach to plagiarism detection (PD) that does not require disclosing any content as cleartext.

As a first step towards this vision, this paper reports on devising a content-protecting variant of the citation-based similarity measure Bibliographic Coupling (BC) \cite{kessler_bibliographic_1963} implemented in our PDS HyPlag \cite{meuschke_hyplag:_2018}. We present an approach to securely mask bibliographic references and an adaption of the Private Set Intersection (PSI) approach to compute the bibliographic coupling strength (BCS) of papers without revealing the cleartext of the references. 

\section{Related Work} \label{sec:relwork}
Our research shall enable plagiarism detection systems to identify similar content in documents without disclosing the content. Data security research has yielded two approaches to protect content while allowing the identification of the cleartext: Revertible functions (encryption) and lossy \mbox{one-way} functions (hashing). 

\textit{Encryption} is conceptually less secure than hashing because the encrypted content contains the full information of the cleartext. A malicious party can gain access to the cleartext by obtaining the decryption key. Additionally, encryption methods considered secure today can become vulnerable in the future due to undiscovered flaws or increases in computing power \cite{chen_report_2016}.   

\textit{Hash functions} are lossy one-way functions that map cleartext of any size to a fixed-sized value (hash). Due to the lossy mapping, the cleartext cannot be recomputed from the hash. The only option for a malicious party to ascertain the correspondence of a hash and the cleartext is to guess the possible cleartext, compute its hash, and compare it to the hashes disclosed for a document. This approach, known as a \textit{preimage attack}, is feasible if the set of hash inputs is finite and known \cite{rogaway_cryptographic_2004}. Privacy-focused messaging applications like Signal face a similar problem called private contact discovery due to the finite set of phone numbers \cite{marlinspike_difficulty_2014}. Signal solved this issue by performing PSI using a secured part of the CPU \cite{maxino_effectiveness_2009}. This solution does not apply to our distributed detection use case, as it still requires trust in the hardware underlying the service.

Researchers predominantly employed hashing to detect document similarity without revealing the documents' content.

The work of Unger et al. \cite{unger_elxa:_2016} is most related to ours since it also protects the content of documents during the plagiarism detection process. Their approach relies on a central authority (root node) that manages the access of peripheral nodes to content in the distributed system. The peripheral nodes represent documents as word chunks, which the nodes process using a collision-resistant hash function with a globally shared salt. The computed hashes are added to a count-min sketch \cite{Cormode2005} for tracking the frequency of word chunks in the document. The count-min sketches are shared with the root node and can be queried by the peripheral nodes. The privacy of all communication within the system is secured using the TLS protocol, with the root node acting as a certificate authority. While the approach greatly improves the confidentiality of content compared to traditional PDS, the count-min sketches are vulnerable to dictionary attacks. Moreover, the root node receives meta-data about documents, which can include the documents' subject matter and information on the authors' writing style.

Murugesan et al. \cite{murugesan_efficient_2010} proposed the use of bloom filters in combination with hashes to protect the semantic meaning of content but maintain knowledge about the content's composition to perform similarity detection tasks. Bloom-filters are conceptually related to count-min sketches. Garbled Boom Filters track the frequency of content chunks in a document using a fixed-sized map while filling the empty positions with noise. 

A drawback that all of the aforementioned hash-based approaches share is their vulnerability to preimage attacks. Furthermore, the approaches rely on a central authority.

\textit{Secure Multi-Party Computation (SMPC)} \cite{goldreich_how_1987} describes methods that overcome the need for a central authority. In SMPC, parties jointly compute a function over inputs, which the parties keep private. Most SMPC protocols, however, require translating all computations to binary circuits \cite{riazi_mpcircuits_2019}. Employing SMPC for plagiarism detection would thus require new implementations of PD methods. 

Many PD tasks represent an exchange of data between two instead of $n$ parties, hence do not require the application of elaborate SMPC protocols. Instead, the tasks can be solved using the \textit{Private Set Intersection (PSI)} \cite{chen_fast_2017} approach. PSI allows two parties to compare private versions of their sets of data without revealing information to third parties. Hashing is the core of PSI. Many PD tasks exclusively require ascertaining the existence of identical features in documents of two parties. Hence, we consider PSI as promising for developing content-protecting versions of PD methods.

\section{Method}
In this paper, we consider bibliographic references as the only content to be compared confidentially. The bibliographic coupling algorithm considers the sets of references in the input and comparison document $R_d$ and $R_d'$ to compute the document similarity score bibliographic coupling strength ($s_\mathrm{BC}$) as
\begin{align}
  s_\mathrm{BC}\left(R_d,R_d'\right) &= \frac{\left|R_d \cap R_d'\right|}{\left|R_d \cup R_d'\right|} = \frac{\left|R_d \cap R_d'\right|}{\left|R_d\right|+\left|R_{d'}\right|-\left|R_d \cap R_d'\right|}.
\end{align}

Employing simple hashing to protect the confidentiality of bibliographic references is prone to preimage attacks as the number of published papers, and hence the number of possible references is finite. An attacker could acquire the metadata of most or all references, pre-compute their hashes and compare the pre-computed hashes of arbitrary references to the hashes of a protected document to deduce the topical context of the document.

To prevent preimage attacks for our use case, we hash combinations of $k$ references instead of single references and only disclose the resulting set of combined hashes to the detection service. 
Without loss of generality, we assume that a preprocessing of references has been completed before forming the subsets. We further assume that the preprocessing step i) eliminated any duplicates in the reference lists of individual documents, ii) disambiguated all references in the collection, iii) stored the disambiguated references as a hashable data structure, and iv) excluded documents that contain less than $k$ references from the similarity computation.

We form all $k$-combinations. For example, if a document contains three references $\{a, b, c \}$ and we seek to form reference subsets of cardinality $k=2$, we would form, e.g., the subsets $\{a,b\}$, $\{a,c\}$, and $\{b,c\}$ but not $\{a,a\}$. Formally, we form the reference subsets
\begin{align}
  \mathcal{P}_k(R_d)=\left\{r\subseteq R_d \big| \left|U\right|=k\right\} 
\end{align}
with cardinality
\(
  \left|\mathcal{P}_k(R_d)\right|= {\left|R_d\right| \choose k}.
\)
Growing the set of possible hashes by a power of $k$ increases the cost of a preimage attack but also the complexity of the detection process.
Moreover, document pairs must contain at least $k$ common references to exhibit a similarity that the content-protecting BC method can detect.

To mask the cleartext of references, we compute the set of hashes
\begin{align}
  H_d=\left\{H(r)\big| r \in \mathcal{P}_k(R_d)\right\}.
\end{align}

Here $H(r)=\sum_{i=1}^k H(r_i)$ denotes the hash function over the subset of references $r\subseteq R_d.$
For the case $k=1,$ $H(r)$ yields the hashes of the individual references, i.e., $H_d=\left\{H\left(r_1\right),H\left(r_2\right),\ldots,H\left(r_n\right)\right\}.$

The detection service performs a private set intersection of the hashes from the input document and the hashes from previously submitted documents $H_d'$ to compute the private BCS as
\begin{align}
  s_\mathrm{PBC}\left(H_d,H_{d'}\right) &= \frac{\left|H_d \cap H_{d'}\right|}{\left|H_d \cup H_{d'}\right|}.
\end{align}

Similarly, one can derive \(s_\mathrm{BC}\left(H_d,H_{d'}\right)\) via
\begin{align}
s_\mathrm{BC}\left(H_d,H_{d'}\right) &=
\frac{\Game_k\left( H_d \cap H_{d'}\right)}
{\Game_k(H_d)+\Game_k(H_{d'})-\Game_k\left(|H_d \cap H_{d'}\right)},
\end{align}
where $\Game_k(j)$ is the numeric solution for $j={\Game_k(j) \choose k}.$
For example, for $k=2$ one can derive $\Game_k(j)=\frac{1}{2}(1+\sqrt{8j+1}).$

After comparing the hashes of an input document to the hashes of the corpus, we retrieve potential sources by ranking all corpus documents in descending order of their maximum $s_\mathrm{PBC}$ and filter for matches exclusively occurring in one document pair.

\section{Experiments}
Our experiments analyze the effectiveness, consumption of computational resources, and resistance to preimage attacks for our content-protecting version of the BC algorithm.

\subsection{Experimental Setup}
We conducted our experiments on a dataset of 105,120 arXiv documents, into which we embedded ten confirmed cases of plagiarism, each consisting of the plagiarized document and one source document. We used the same dataset in previous work \cite{meuschke_improving_2019}. We excluded documents without processable reference data and documents with more than 150 references. The final dataset contained 92,082 documents and 1,726,359 unique bibliographic references.

In a preprocessing step, we used the open-source software GROBID\footnote{\url{https://github.com/kermitt2/grobid}} to convert all documents into the uniform TEI-format\footnote{\url{https://tei-c.org/}}. TEI employs XML to structure the documents' content and allows for easy extraction of the bibliographic references.

Initial tests showed that the title field has the highest probability of being present in the reference string. Therefore, we used the normalized title field for the hashing. 
To decide on a hash function to use, we counted the number of hash collisions resulting from applying Adler32, SHA1, and SHA256 for hashing all reference subsets of size 3 in 5,000 documents. Only Adler32 yielded hash collisions (1,320), i.e., mappings of different inputs to the same hash, due to the comparably smaller size of its hashes (32-bits). We thus chose SHA1 as it offers sufficient collision resistance ($2^{80}$) and is faster to compute than SHA256. Collisions would cause false positives and,  thus, an unnecessary effort for human reviewers.

\subsection{Results}
\textbf{Effectiveness.} 
To compare the effectiveness of PBC to the original BC method, we computed $s_{PBC}$ and $s_{BC}$ for all ten test cases in our dataset using subset sizes of 1, 2, and 3, respectively. We found that for both $k=2$ and $k=3$, the similarity scores computed by PBC and BC were equal for all test cases. This result shows that PBC detects identical references equally well as BC.    

\textbf{Resource Consumption.} To assess the computational effort of our content-protecting PBC method, we analyzed the computation time and storage required for computing $s_{PBC}$ depending on the size of $k$. We divided the analysis into two steps. 

In the first step, we assessed the time and storage required for computing and storing the hashed reference subsets. \Cref{tab:100K} shows the results for analyzing the entire dataset of 92,082 documents using different sizes of $k$. The results show that the exponential growth of the space required for storing hashed reference subsets is the limiting factor for using larger subset sizes. 

\begin{table}[]
    \caption{Hash generation for 92,082 documents.} \label{tab:100K}
    \begin{tabular}{llcc}
    \textbf{k-Tuple} & \textbf{Hashes} & \textbf{Time in sec} & \textbf{Size in GB} \\
    \midrule
    k = 1            & 1,726,359        & 21        & 0.185 \\
    k = 2            & 45,951,328       & 31        & 2.5   \\    
    k = 3            & 1,848,313,500    & 258       & 126   \\           
    \bottomrule
  \end{tabular}
\end{table}

In the second step, we assessed the time required for performing the private set intersection of the hashed reference subsets depending on $k$. For this analysis, we only used 1,000 documents from our dataset. \Cref{tab:1K} shows the number of all hashes and the fraction of those hashes that occur in one, two, and three documents, respectively. The table also shows the time required for applying PBC to compute the bibliographic coupling strength of the input document with a comparison document. For increasing values of $k$, the number of hashes occurring in more than one document decays rapidly. For $k=1$, $86\%$ of the references occur in one document only. Combinations of three references are unique in $99.3\%$ of the cases. The increase in required computation time is almost constant in the number of hashes due to the use of indexes.

\begin{table}[]
    \caption{Detection against 1,000 documents} \label{tab:1K}
    \begin{tabular}{lrrr}
    \textbf{k-Tuple} & \textbf{Hashes} & \textbf{Ratio in 1/2/3 docs} &  \textbf{Time in ms} \\
    \midrule
    k = 1            & 22,658           & .86/.07/.03           & 98 \\
    k = 2            & 357,765          & .98/.02/.001          & 103  \\
    k = 3             & 5,250,076       & .99/.01/.0            & 118  \\           
    \bottomrule
  \end{tabular}
\end{table}

\textbf{Resistance to Preimage Attacks.}
To motivate the resistance of PBC to preimage attacks, we estimate the computation time required to perform such an attack, using computer science publications as an example. As explained in \Cref{sec:relwork}, a preimage attack requires knowing the possible hash inputs, i.e., in our case, the possible references in academic documents. To estimate the number of possible references for computer science, we use the dblp bibliography\footnote{\url{https://dblp.uni-trier.de}}, which is the most comprehensive collection of bibliographic records for journal articles, conference papers, and monographs in this field. As of May 2020, dblp contains 5.05 million records.  

For $k=1$, a preimage attack on a single computer science document requires computing \({{5.05\times 10^6}\choose{1}} = {5.05\times 10^6}\) hashes, i.e., a time complexity of $\mathcal{O}(n^k)$. Assuming a computing time of 1ms per hash would result in a single-threaded runtime of $5.05\times 10^3s\approx1.40\mathrm{h}$.
Analogously, for $k=2$, computing \({{5.05\times 10^6}\choose{2}} \approx 12.75\times 10^{12}\) hashes requires a single-threaded runtime of more than 404 years and more than 680 million years for $k=3$. We see these numbers as conservative lower-bound estimates of the effort required because our calculation ignores i) interdisciplinary citations, ii) citations to sources not formally published, such as websites, code libraries, and datasets, and iii) the inevitable incompleteness of dblp\footnote{\url{https://dblp.uni-trier.de/faq/23593238.html}}. 

The time complexity of $\mathcal{O}(n^k)$ for performing a preimage attack on a single document makes this attack too costly for $k\geq2$ if the attacker cannot limit the possible references significantly, e.g., to a narrow research field. For example, for $k=2$ and $x=$ 30,000 possible references, the attack requires a single-threaded runtime of approx. $125\mathrm{h}$ for one document. For $k=3$, the required single-threaded runtime exceeds $100\mathrm{h}$ per document for $x\geq$ 1,300. We hypothesize that being able to restrict the possible references that sharply reduces the novel, i.e., unexpected information an attacker could obtain, hence reduces the benefit of a preimage attack greatly.

\section{Conclusion and Future Work}
We proposed PBC - a method that computes the bibliographic coupling strength of documents without revealing the bibliographic references involved in the computation. To realize PBC, we invented a new PSI approach that uses hashed feature subsets to prevent preimage and dictionary attacks. The security of the approach is adjustable to the computing resources that might be spent on the specific problem by increasing the number of features included in a subset, and by using hash functions with higher bit-lengths. 

We showed that PBC using SHA1 capably identifies similar documents in a large corpus without causing hash collisions. We demonstrated that a subset size of $k=2$ achieves the best trade-off between computation time, required storage, and attack resistance. A subset size of $k>2$ causes a steep rise in computation time and is therefore limited to documents with small numbers of references. 

Hashed document feature subsets show promise for building a future decentralized PD service since accuracy would only decrease if a document pair does not contain at least two matching references. As shown in our prior work \cite{GippMB14}, an overlap of two references generally does not constitute a similarity that is significant enough to identify a document as suspicious of plagiarism.

In the future, we will allow encrypting the document IDs of unpublished documents.
By doing so, the PDS can still detect that an input document overlaps with already existing unpublished documents in the distributed reference database. However, the PDS can no longer determine the number of documents with which the input document shares content. By using incentive mechanisms, the owner of the unpublished document is motivated to share the document with the PDS privately. To avoid false positives, authors can, e.g., submit previous versions of rejected grant proposals to prove that they were the authors of the earlier document.

In this initial study, we focused entirely on Bibliographic Coupling and excluded more sophisticated citation-based plagiarism detection methods like Greedy Citation Tiling and Longest Common Citation Sequences \cite{GippM11}. In the future, we will devise content protection methods that support pattern-based PD approaches that use mathematical features \cite{meuschke_improving_2019} and images \cite{meuschke_adaptive_2018}. We will analyze these detection methods and examine which features need to be masked and which features can be shared openly during the detection process without revealing any semantic information. 

In summary, we successfully conducted the first step towards our vision of a decentralized content protecting plagiarism detection system \cite{ihle_privacy-preserving_2019}. The findings of our initial study confirm our research direction of using hashed document features, such as references, in-text citations, images, and formulae, to devise such a service.

To ensure the reproducibility of our experiments, our data and code are available at \textit{\url{https://github.com/ag-gipp/20CppdData}}\\

\printbibliography[keyword=primary]

%% file: references.bib
@InProceedings{Ihle2020,
  title = {A {First} {Step} {Towards} {Content} {Protecting} {Plagiarism} {Detection}},
  booktitle = {Proceedings of the {ACM}/{IEEE} {Joint} {Conference} on {Digital} {Libraries} ({JCDL})},
  author = {Ihle, Cornelius and Schubotz, Moritz and Meuschke, Norman and Gipp, Bela},
  year = {2020},
  month = {Aug.},
  topic = {pd},
  doi = {10.1145/3383583.3398620},
}
